\documentclass[aps,pre,twocolumn,superscriptaddress]{revtex4}
\usepackage{amsfonts}
\usepackage{amsmath}
\usepackage{amssymb}
\usepackage{graphicx}

\begin{document}

\title{ Induced cooperative motions in a medium driven at the nanoscale: Searching for an optimum excitation period }


\author{V. Teboul}
\email{victor.teboul@univ-angers.fr}
\affiliation{Laboratoire de Photonique d'Angers EA 4464, Universit\' e d'Angers, Physics Department, 2 Bd Lavoisier, 49045 Angers, France}

\author{J.B. Accary}
\affiliation{Laboratoire de Photonique d'Angers EA 4464, Universit\' e d'Angers, Physics Department, 2 Bd Lavoisier, 49045 Angers, France}

\keywords{dynamic heterogeneity,azobenzene,isomerization,glass-transition}
\pacs{64.70.pj, 61.20.Lc, 66.30.hh}

\begin{abstract}
Recent results have shown the appearance of induced cooperative motions called dynamic heterogeneity during the isomerization of diluted azobenzene molecules in a host glass-former.  In this paper we raise the issue of the coupling between these "artificial" heterogeneities and the isomerization period.
How do these induced heterogeneities differ in the saturation regime and in the linear response regime ?
Is there a maximum of the heterogeneous motion versus isomerization rate  and  why ? 
Are the heterogeneity evolution with the isomerization rate connected with the diffusion or relaxation time evolution ?
We use out of equilibrium molecular dynamics simulations to answer these questions.
We find that the heterogeneity increases in the linear response regime for large isomerization periods and small perturbations.
In contrast the heterogeneity decreases in the saturation regime, i.e. when the isomerization half-period ($\tau_{p}/2$) is smaller than the relaxation time of the material ($\tau_{\alpha}$). 
This result makes possible a test of the effect of cooperative motions on the dynamics using the chromophores as Maxwell demons that destroy or stimulate the cooperative motions.
Because the heterogeneities increase in the linear regime and then decrease in the saturation regime, we find a maximum for $\tau_{p}/2 \approx \tau_{\alpha}$.
The induced excitations concentration follows a power law evolution versus the isomerization rate and then saturates. As a consequence the $\alpha$ relaxation time is related to the excitation concentration with a power law, a result in qualitative agreement with recent findings in constrained models.
This result supports a common origin for the heterogeneities with constrained models and a similar relation to the excitation concentration.\\
{\em Published in Phys. Rev. E {\bf 89}, 012303 (2014).}\\
\end{abstract}

\maketitle
\section{Introduction}

Since their discovery  dynamical heterogeneities\cite{h0} (DHs) have attracted much interest as a possible route towards the explanation of the glass-transition mechanism\cite{binder,bene}. Indeed, the presence of these heterogeneities has already successfully explained the Stokes-Einstein breaking\cite{se} mechanism in glass-formers, the non-Gaussian evolution of the Van-Hove correlation functions and at least a significant part of the stretching of the density autocorrelation functions or incoherent scattering functions. 
Dynamical heterogeneities are also a key feature in several models of the glass transition (constraint models\cite{ha,hb,hc}, Adam-Gibbs configurational entropy model, ...). 
Recently it has been found that the heterogeneities can be controlled  using the isomerization of a few azobenzene molecules dispersed inside the glass-former\cite{prl,pre}. 
The origin of these induced heterogeneities may be found in the Onsager regression hypothesis for small perturbations\cite{david}.
If  small external perturbations behave as the spontaneous fluctuations of the medium, then DHs can be created from small perturbations, here the azobenzene isomerizations, in the same way than spontaneous fluctuations create the spontaneous  heterogeneities.
Note however that the way in which spontaneous fluctuations create the DHs is still not understood. 
Despite this fact the exciting possibility to control the DHs opens a route to use this control as a new tool to test the relation between the heterogeneities and the transport properties of glass-formers, or  even to use the DHs to control the transport properties of the material.
Indeed an acceleration of the dynamics has been reported, from simulations\cite{prl} and experiments\cite{nature,natmat}, in association with the induced DHs.
However the different parameters that control the strength of the induced heterogeneities as well as their characteristic time, usually named $t^{*}$, have first to be determined carefully. 
In this paper we thus raise the issue of the relation between the isomerization period and the induced dynamical heterogeneities.
In particular we expect interesting couplings when the isomerization half-period $\tau_{p}/2$ will be equal to the relaxation time $\tau_{\alpha}$ of the host material.
We find that the heterogeneities increase with $1/\tau_{p}$ when the isomerization half-period is larger than $\tau_{\alpha}$ and then decrease for smaller periods.
The DHs are thus maximal for $\tau_{p}/2 \approx \tau_{\alpha}$.
Note that this maximum of the heterogeneities occurs around the same period than the maximum of the diffusion, i.e. at the end of the linear diffusion regime.

\section{Calculations}

We simulate the photoisomerization of one dispersed red ($DR1$) molecule ($C_{16}H_{18}N_{4}O_{3}$, the probe) inside a matrix of 500 methylmethacrylate ($MMA$) molecules ($C_{5}H_{8}O_{2}$, the host). A detailed description of the simulation procedure can be found in previous works\cite{prl,probe,cage,jcp}. The main differences are that:
i) The concentration of probe molecules is here much smaller and the size of the box larger.
ii) To speed up the calculations, we model the host interactions here with a recent coarse-grain potential function\cite{pot2}. For the probe molecule, we use however the same all-atom site-to-site interaction potential\cite{pot1} than previously.
The density is set constant at $1.19 g/cm^{3}$.
We thus have 7541 atoms in a 41.26 \AA\ wide cubic simulation box. 
We use the Gear algorithm with the quaternion method \cite{allen} to solve the equations of motions with a $\Delta t=10^{-15} s$ time step . The temperature is controlled using a Berendsen thermostat \cite{beren}.
We model the isomerization as a uniform closing and opening of the probe molecule shape\cite{fermeture,mecanisme1,mecanisme2,mecanisme3} during a characteristic time $t_{0}=1ps$; while the period of the isomerization cis-trans and then trans-cis is a variable that we refer as $\tau_{p}$ in the article. As a result $\tau_{p}/2$ appears as the time lapse between two energetic impulses inside the material and will be an important characteristic time in our study.
In our simulations  the isomerization takes place at periodic intervals whatever the surrounding local viscosity.
This main approximation has recently been validated experimentally\cite{barrettn,nature}, as the viscosity necessary to stop the isomerization of this probe molecule is extremely large\cite{barrettn}.
In order to evaluate the possible aging of our results we make two successive simulations of $10$ ns  each for every data presented in this paper.
We then compare the results for the two different simulations. Surprisingly enough the results show only slight differences in the saturation regime.
However in our Figures we display the most aged data or occasionally both data.
Finally, in this work we use a small concentration of chromophores. There is only one chromophore inside the simulation box.  
We use this small concentration to have, in our simulations, the smallest possible perturbations of the host material for a given isomerization rate. 
The main drawback due to this small concentration is that the observed effects are smaller. 
\\

\section{Results and discussion}
\subsection{1. What is expected about the induced heterogeneity dependence on the  isomerization period $\tau_{p}$ ?}

For very large isomerization periods $\tau_{p}$, we expect the medium to stay unchanged or be only slightly perturbed. We expect this regime to last while $\tau_{p}/2 > \tau_{\alpha}$, as when this condition is verified the time between two isomerizations is larger than the relaxation time of the material. As a result the material is in this case relaxed before that a new isomerization takes place and the medium perturbations are due to one isomerization only.
In a previous study \cite{jcpvic} we found that the diffusion coefficient as well as the inverse of the $\alpha$ relaxation time increase linearly with the frequency of isomerizations $f=1/\tau_{p}$ in that regime. According to this result  we will name that regime, {\em the linear regime} as in \cite{jcpvic}. Note that the term linear refers here to the diffusion coefficient evolution.
In the linear regime we expect the isomerization-induced cooperative motions (DHs) to increase with the frequency $f$ due to the increase of the number of pulses, as each pulse has the same mean probability to initiate a cooperative motion in a relaxed material. 
Note that this expectation will be true only if the isomerizations stimulate the DHs creation, else the cooperative motions will decrease when increasing $f$ as each action on the medium leads eventually to an increase of the medium effective temperature. 
For larger frequencies we expect finally the cooperative motions ever to saturate or to decrease due to the host softening.
We will name that regime {\em the saturation regime} referring to the saturation of the diffusion that appears for large frequencies\cite{jcpvic}.

\vskip 1cm
\subsection{2. How does the heterogeneity differ in the saturation regime and in the linear response regime ?}
In order to now investigate the coupling between the isomerization period $\tau_{p}$  and the isomerization-induced dynamic heterogeneities,  we display in Figures 1 and 2 for various isomerization rates, two different quantities that measure the DHs, namely the non-Gaussian parameter and the four-points dynamic susceptibility.

The Non-gaussian parameter $\alpha_{2}(t)$ is defined as:

\begin{equation}
\alpha_{2}(t)=\frac{3 <r^{4}(t)>} {5<r^{2}(t)>^{2}}    -1  \label{e4}
\end{equation}

And we used the following definition for the dynamic susceptibility $\chi _{4}$\cite{h0}:

\begin{equation}
\chi _{4}(a,t)=\frac{\beta V}{N^{2}}\left( \left\langle
C_{a}(t)^{2}\right\rangle -\left\langle C_{a}(t)\right\rangle ^{2}\right) 
\label{e2}
\end{equation}
with 
\begin{equation}
C_{a}(t)={\sum_{i=1}^{N}{w_{a}}}\left( \left\vert {{{\mathbf{r}}}}_{i}(t)-{{{\mathbf{r}}}}_{i}(0)\right\vert \right) .  \label{e3}
\end{equation}
Here, $V$ denotes the volume of the simulation box, $N$ the number of molecules in the box, and $\beta =(k_{B}T)^{-1}$. The symbol $w_{a}$ stands for a "discrete mobility"  window function, $w_{a}(r)$, taking the values $w_{a}(r)=1$ for $r>a$ and zero otherwise. 
In our calculations we have used the parameter a=$1.0$ \AA\ that was found previously\cite{pre} to lead to the larger $\chi_{4}$ in our system.

Our simulation box is a cube, $41.26$ \AA\ wide. A radius $R=20$ \AA\ around the chromophore includes thus approximately the whole simulation box.
However because when the distance $R$ from the chromophore increases the isomerization effects progressively disappear, we will focus our attention on the molecules situated within a radius $R=10$ \AA\ from the chromophore.
\includegraphics[scale=0.34]{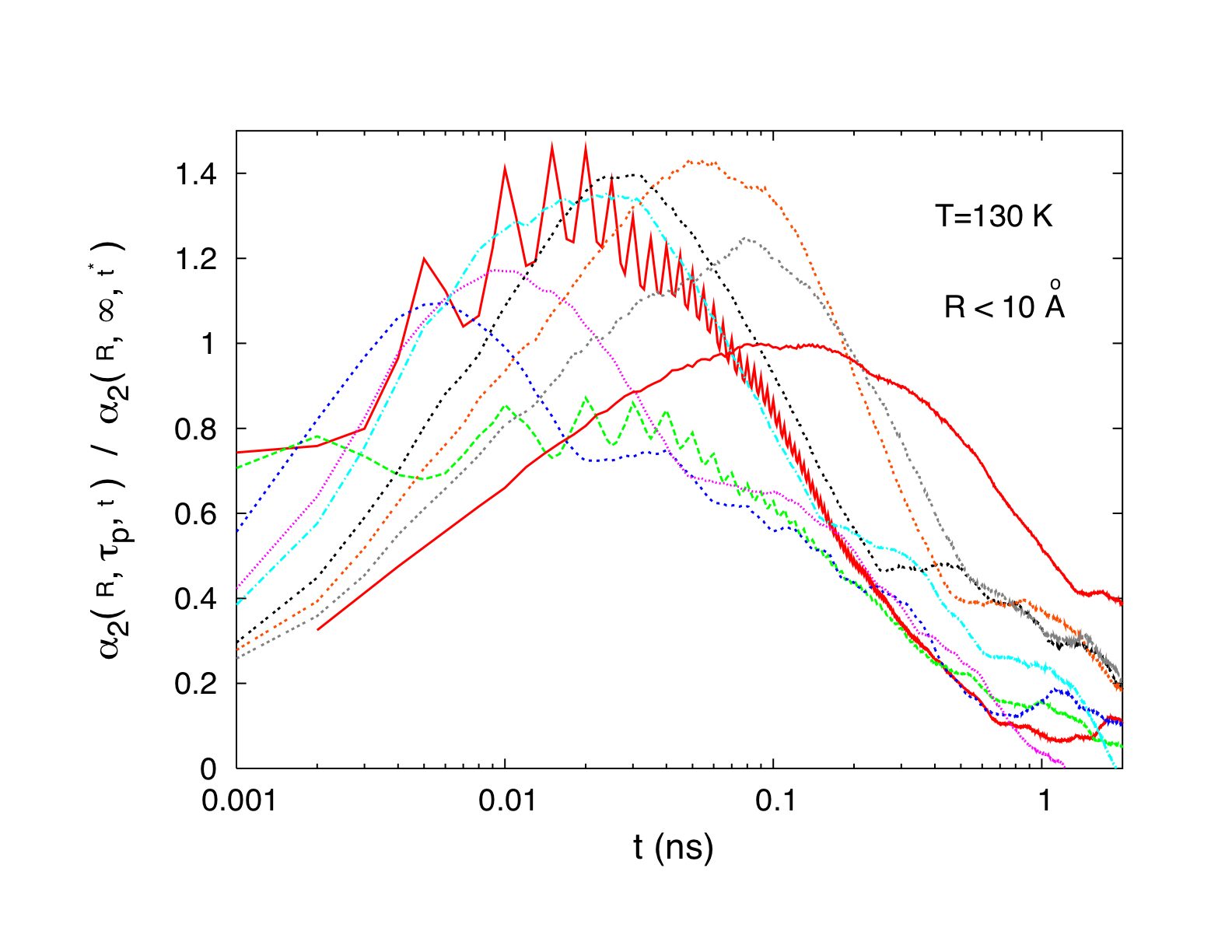}
{\em \footnotesize Figure 1:
Non-Gaussian parameter (NGP) $\alpha_{2}(R,\tau_{p},t)$ for various isomerization rates $f=1/\tau_{p}$ and for molecules situated at a distance $R<10$\AA $ $ from the chromophore.  $\alpha_{2}$ measures the divergence of the Van Hove correlation function from the Gaussian behavior expected for a Markovian process. In supercooled liquids this non-Gaussian behavior of the Van Hove arises from the appearance of a tail at large $r$ induced by the cooperative motions of the most mobile molecules (i.e. DHs).  As a result $\alpha_{2}$ measures the DHs. In the Figure $\alpha_{2}(R,\tau_{p},t)$ is divided by its maximum value for  $\tau_{p}=\infty$ and has no unit. Values larger than $1$ mean that the isomerizations increase the Non-Gaussian Parameter. In most cases this result means an increase of the DHs. For the shorter periods $\tau_{p}$ displayed in the Figure however, $\alpha_{2}$ can also be increased by the forced motions induced by the isomerizations. 
From the left to the right: 
green dashed curve: $\tau_{p}=10 ps$, blue dashed curve:  $\tau_{p}=50 ps$, pink dotted curve: $\tau_{p}=100 ps$, blue dotted dashed curve: $\tau_{p}=200 ps$, black dotted curve: $\tau_{p}=500 ps$, orange dotted curve: $\tau_{p}=1000 ps$, grey dotted curve: $\tau_{p}=2000 ps$, red continuous line:  
$\tau_{p}=\infty$.
The other continuous (red) line corresponds to $\tau_{p}=5 ps$.
The maximum value of the different $\alpha_{2}$ is observed for $\tau_{p}=1000 ps$.}
 \vskip 0.5cm
Figure 1 shows the Non Gaussian Parameter (NGP) $\alpha_{2}(t)$ for various isomerization periods ranging from $\tau_{p}=5$ $ps$ to $\tau_{p}=\infty$. For the smaller period considered ($\tau_{p}=5$ ps) we observe a large NGP, and peaks located at the isomerizations periods $t=n.\tau_{p}$ with integer values of $n$. Increasing the isomerization period $\tau_{p}$ we observe a strong decrease for $\tau_{p}=10$ ps, while the peaks are still visible. Increasing again the period we observe a continuous increase of $\alpha_{2}(t)$ together with an increase of the NGP characteristic time $t^{*}$ while the peaks are progressively washed out due to their widening.
The Non Gaussian Parameter reaches a maximum for $\tau_{p}=1$ ns and then  begins to decrease for larger periods, down to its value without isomerization.


\includegraphics[scale=0.34]{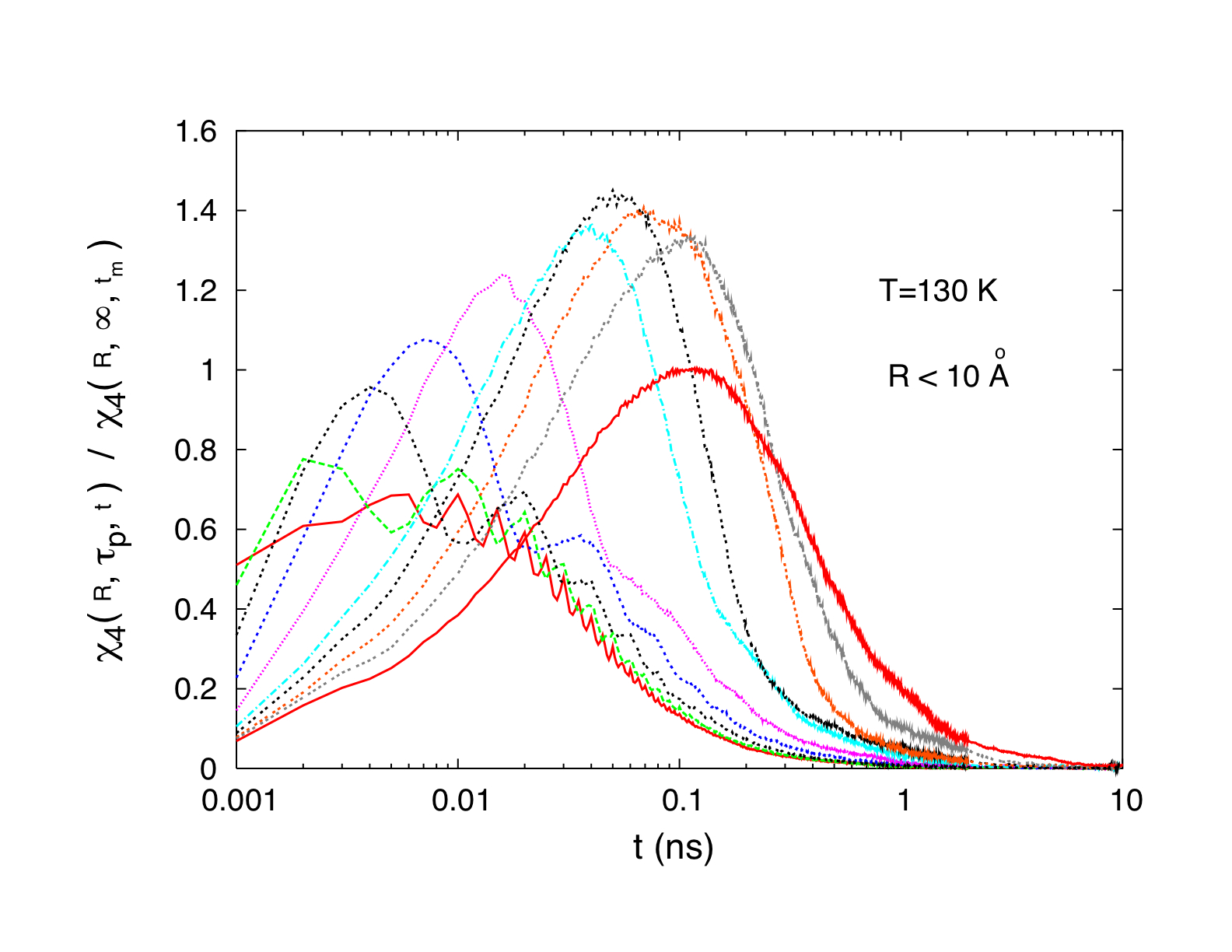}
{\em \footnotesize  Figure 2a:
4-th order dynamic susceptibility $\chi_{4}(R,\tau_{p},t)$ for various isomerization rates $f=1/\tau_{p}$ and for molecules situated at a distance $R<10$\AA $ $ from the chromophore.
$\chi_{4}(R,\tau_{p},t)$ measures the variance of the fluctuations of the mobility, and as a result the DHs.
The susceptibility $\chi_{4}(R,\tau_{p},t)$ is divided by its maximum for $\tau_{p}=\infty$ with the same value of $R$ and has no unit. Values larger than $1$ mean that the isomerizations increase the dynamic heterogeneities.
From the left to the right: 
continuous (red) line corresponds to $\tau_{p}=5 ps$, green dashed curve: $\tau_{p}=10 ps$, blue dashed curve:  $\tau_{p}=50 ps$, pink dotted curve: $\tau_{p}=100 ps$, blue dotted dashed curve: $\tau_{p}=200 ps$, black dotted curve: $\tau_{p}=500 ps$, orange dotted curve: $\tau_{p}=1000 ps$, grey dotted curve: $\tau_{p}=2000 ps$, red continuous line:  
$\tau_{p}=\infty$.
The maximum value of the different susceptibilities is observed for $\tau_{p}=500 ps$.}

\vskip 0.5cm
Figure 2a shows the dynamic susceptibility evolution with the isomerization period.
In contrast with $\alpha_{2}(t)$ in Figure 1, the smaller period $\tau_{p}=5$ ps doesn't lead to a particular  $\chi_{4}(R,\tau_{p},t)$ value, but follows the same trend than the other periods.
We conclude that the particularly large value of $\alpha_{2}(t)$ observed in Figure 1 for $\tau_{p}=5$ ps is not due to an increase of cooperative motions that will result in a similar increase of the susceptibility, but to an increase of perturbed molecular motions that results from a large isomerization rate $1/\tau_{p}$.
If we except this period ($\tau_{p}=5$ ps) we see a very similar behavior for the susceptibility in Figure 2 and the Non Gaussian Parameter in Figure 1.
For large isomerization rates (i.e. small periods) the heterogeneity is smaller than without isomerization.
For example for $\tau_{p}=5$ ps the susceptibility reaches only 60 percents of its value without isomerization.
We thus have a localized decrease of the heterogeneity for small periods (i.e. in the saturation regime).
Then the susceptibility increases progressively with the period and the characteristic time of the susceptibility $t_{m}$ also increases.
When $\tau_{p}$ reaches $500$ ps the susceptibility is maximum and begins to decrease progressively to the $\tau_{p}=\infty$ (i.e. iso. off) value.
Note that the susceptibility reaches a maximum for $\tau_{p}=500$ ps while the NGP reaches a maximum for $\tau_{p}=1000$ ps.
However the two functions are approximately constant around their maximum i.e. between these two $\tau_{p}$ values.
We also note that the two functions increase similarly up to around 140 percent of their spontaneous value.

\includegraphics[scale=0.34]{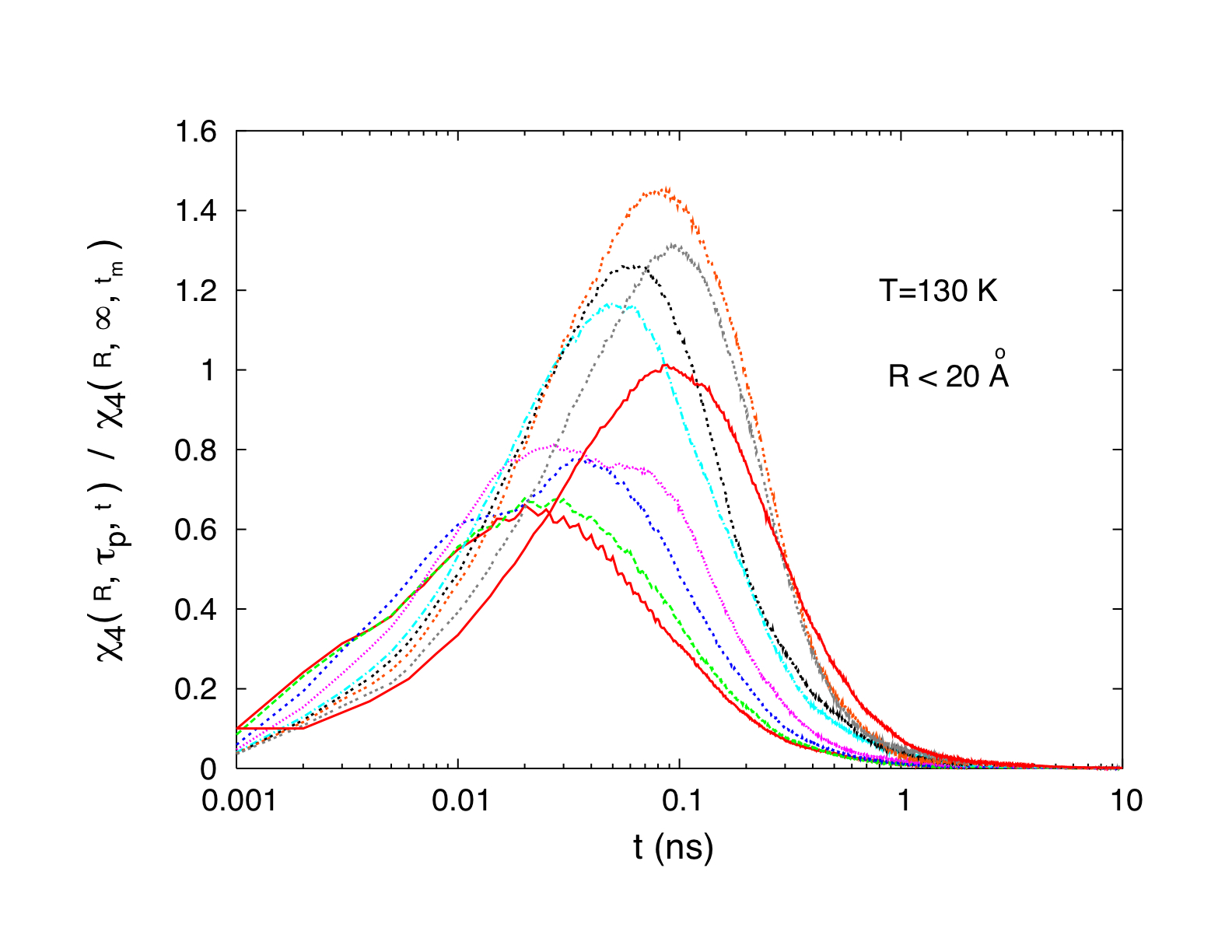}
{\em \footnotesize Figure 2b:
As in Figure 2a but for molecules situated at a distance $R<20$\AA $ $ from the chromophore.}
\vskip 0.5cm
Figure 2b shows the dynamic susceptibility evolution with the period, but with a larger radius than Figure 2a.
The susceptibilities are here calculated including most molecules of the simulation box.
We see however in Figure 2b a similar comportment than in Figure 2a.
The susceptibility first increase with the period, up to 140 percents of the spontaneous susceptibility maximum value.
Then it decreases progressively to the spontaneous susceptibility.
The main differences are that: i) the characteristic times distribution is smaller here.
ii) the maximum occurs for $\tau_{p}=1000$ ps instead of $500$ ps.

\includegraphics[scale=0.34]{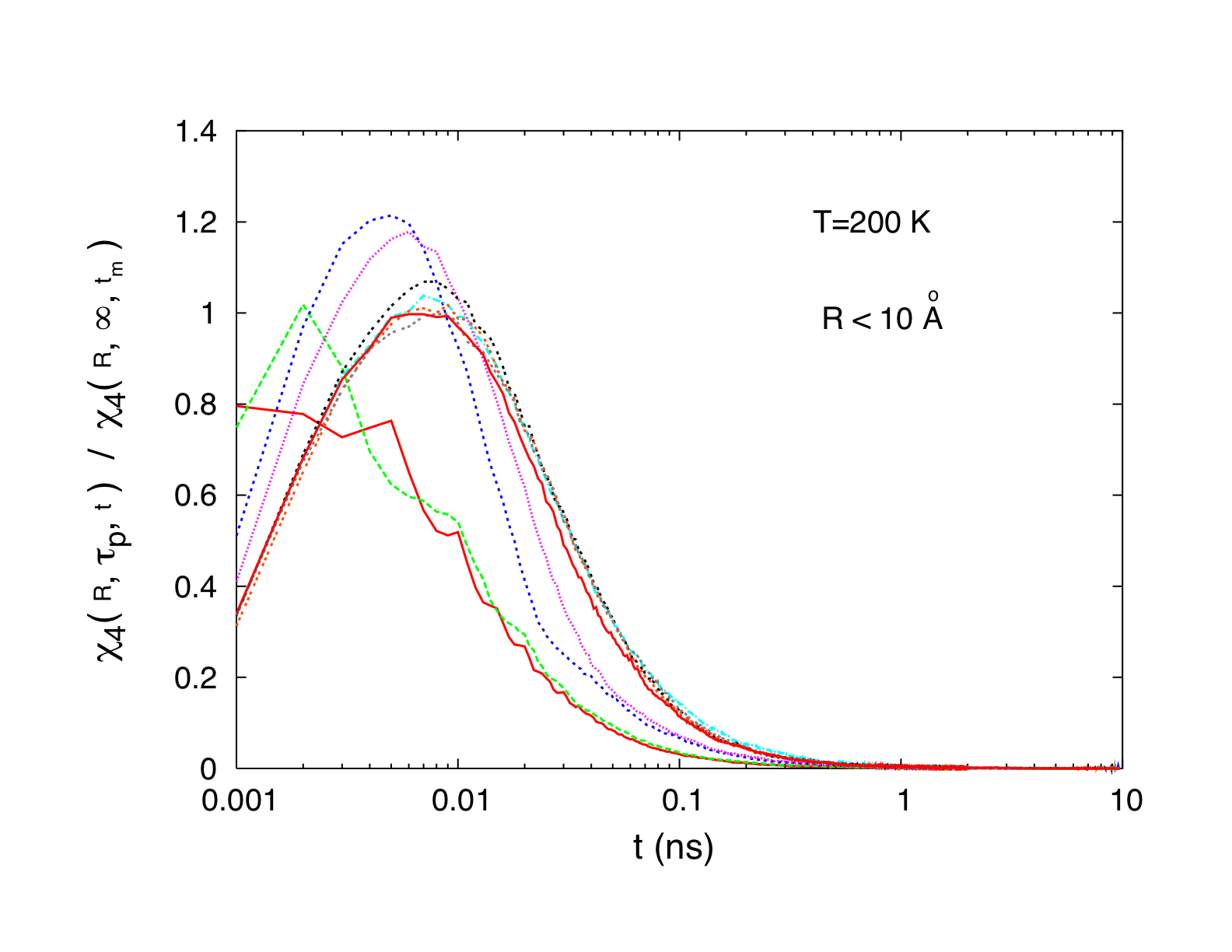}
{\em \footnotesize Figure 2c:
As in Figure 2a but at a larger temperature $T=200K$.}
\vskip 0.5cm
Figure 2c shows at a larger temperature (T=200K), the dynamic susceptibility evolution with the period.
Again we see the same comportment.
The characteristic times are however smaller and a maximum of $\chi_{4}$ is reached for $\tau_{p}=100$ ps instead of $500$ ps.
Note that for $\tau_{p} \gtrsim 500$ ps all curves collapse to the iso off $\tau_{p}=\infty$ susceptibility, showing that the induced DHs disappear for large periods.

\vskip 1cm
To summarize, Figures 1 and 2 show, for distances $R<10$ \AA\ from the isomerizing chromophore, first an increase of the DHs with $f$ and then a rapid decrease.
These increase and rapid decrease appear respectively in the time ranges of the linear regime and of the saturation regime.  These two opposite evolutions of the heterogeneities explain why, testing glass formers under various solicitations, a number of works have found a decrease of the heterogeneities\cite{h5,h6,h6b,h7} while a few others found an increase of the DHs\cite{h8,prl,pre}. Note that the increase of these cooperative motions appears for small solicitations only, and is thus much more difficult to observe experimentally than the opposite evolution.
The increase of the DHs with $f$  shows that the isomerizations stimulate the cooperative motions in the linear regime. This result confirms the conclusion of ref.\cite{prl,pre}.

In contrast, in the saturation regime the DHs decrease sharply around the chromophore. 
This decrease of the heterogeneities for large frequencies below the level of spontaneous DHs, means that we can also destroy the heterogeneities using the isomerizations. This is an interesting result as it makes possible a test of the DHs effect on the dynamics by destroying the heterogeneities when and where they appear, using the chromophores as Maxwell demons. Note that a similar test is also possible from a stimulation of the heterogeneities in the linear regime.

For frequencies $f$ in between the two regimes ($\tau_{p}/2 \approx \tau_{\alpha}$), the medium is still unaffected but we expect non linear contributions due to the interactions between excitations (i.e. the isomerizations induced fluctuations).
These nonlinearities contain informations about the facilitation mechanism, because the facilitation can be seen as an interaction mechanism between excitations.

A previous study has shown\cite{jcpvic} that the saturation regime is related to the appearance of increasingly soft regions around the chromophores surrounded by harder regions. However soft regions, characterized by their smaller relaxation times, have smaller DHs, because the heterogeneities follow the relaxation time of the medium.
Thus we interpret the decrease  of the heterogeneities in the saturation regime as arising from the increased softening around the chromophores.
In another similar picture the softening is equivalent to a local increase of the effective temperature of the host material that will also lead to a decrease of the heterogeneities.
This comportment may change however at larger radii from the chromophore, as the heterogeneity of the structure may lead to a local increase of DHs at the limit between soft and hard zones.
In contrast in the linear response regime the host medium is roughly not affected and thus each solicitation has the same probability to lead to cooperative motions.
As a result the DHs increase with the number of solicitations per second (i.e. with $f$).
Due to this increase of the DHs followed by a decrease, we observe a maximum of the DHs for a characteristic value of the period $\tau_{p}$ that we will name $\tau_{p}^{*}$.
Figures 2  show that $\tau_{p}^{*}$ increases when the temperature drops following the relaxation time of the host material.
$t^{*}$ becomes also larger when the period $\tau_{p}$ increases (i.e. $f$ decreases).
This maximum is situated around the relaxation time of the host material ($\tau_{p}^{*}/2 \approx \tau_{\alpha}$). 
This relation results in an increase of $\tau_{p}^{*}$ when the temperature drops.

\vskip 1cm
\subsection{3. What is the range of stimulation of the heterogeneities ? }
We will now investigate the evolution of the DHs as a function of the distance from the chromophore periodic perturbation.
Our aim here is to test our assumption of different zones in the saturation regime that appear due to the material softening around the chromophore.
Figure 3a shows that, in the saturation regime,  the first peak of $\chi_{4}(t)$ decreases rapidly when we move away from the chromophore.
\includegraphics[scale=0.34]{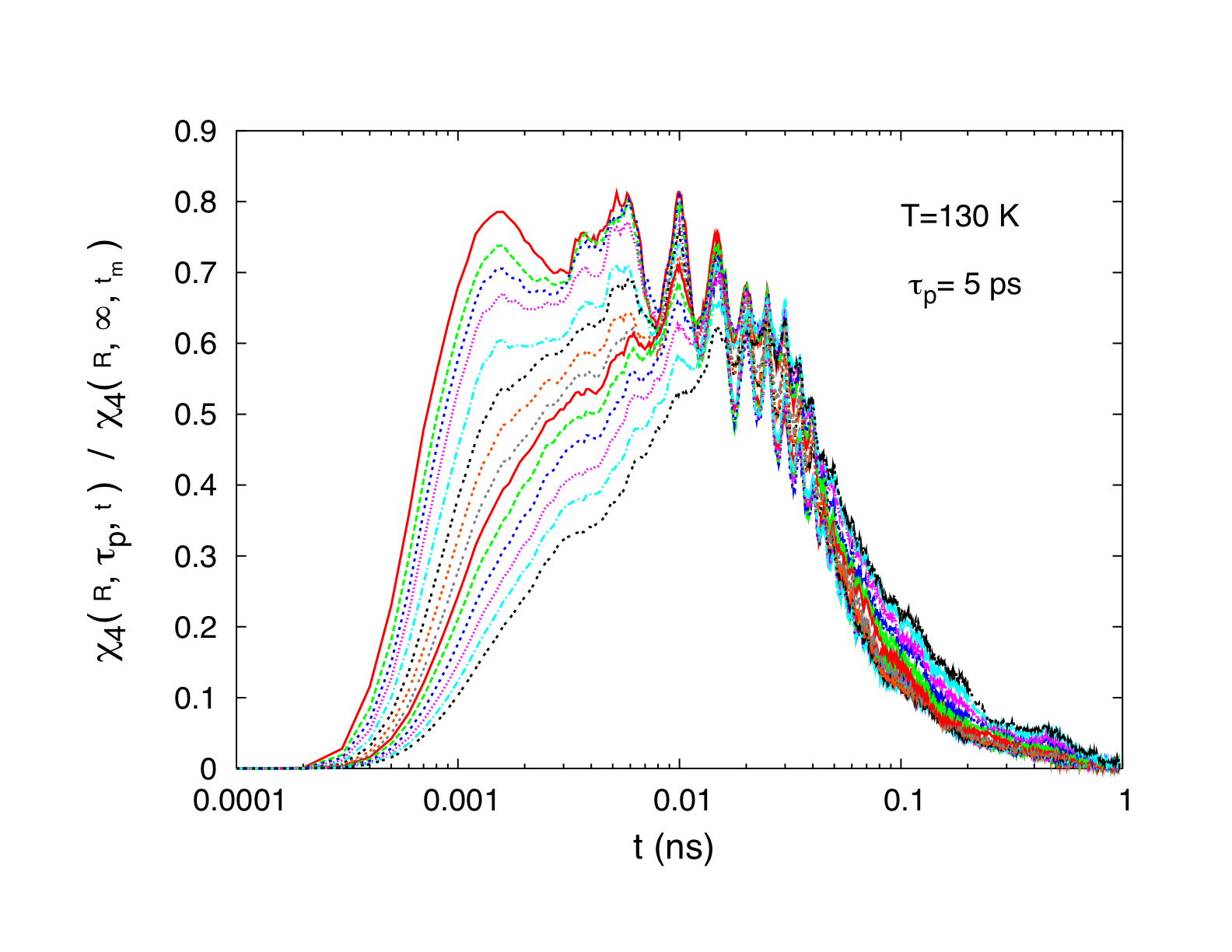}
{\em \footnotesize Figure 3a:
4-th order dynamic susceptibility $\chi_{4}(R,\tau_{p},t)$ for various distances $R$ from the chromophore.
The isomerization rate is chosen in the saturation regime $\tau_{p}=5 ps$.
From top to bottom,  the distances increase by steps of $1$ \AA\ from $0<R<7$ \AA\ (continuous red line) to $0<R<20$ \AA\ (black dashed line).
 \\}

The characteristic times of the different peaks show that the first peak of $\chi_{4}(t)$ corresponds to the DHs induced by one isomerization only while the following peaks correspond to DHs induced by multiples isomerizations. 
The decrease of the DHs then accelerates for $R > 11$ \AA. 
However while the first peak decreases rapidly when R increases, the other peaks stay roughly constant.
This result shows that the effect of one isomerization is spatially limited to a radius $R \approx 11$ \AA\ around the chromophore and that successive isomerizations are necessary to induce DHs at larger distances.
\includegraphics[scale=0.34]{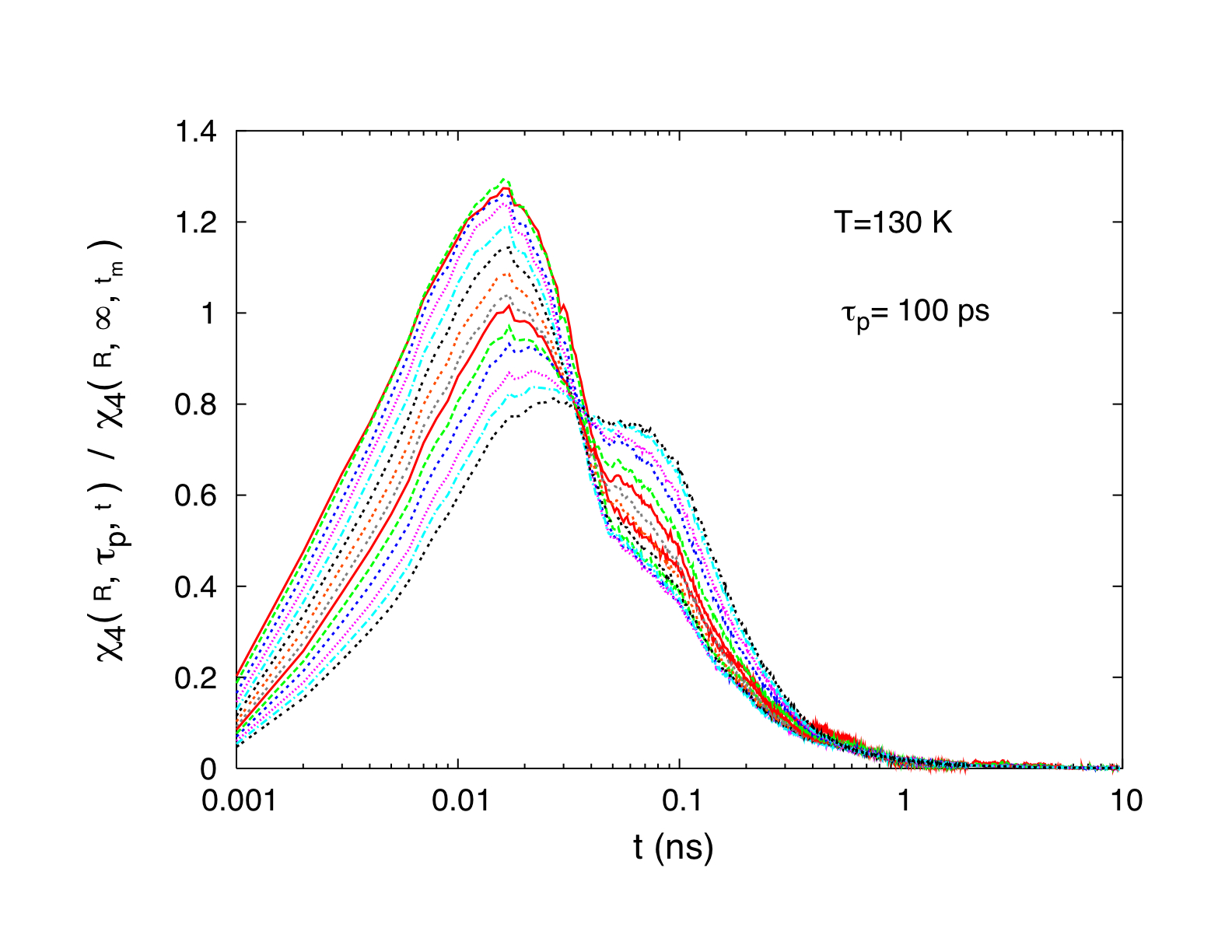}
{\em \footnotesize Figure 3b:
As in Figure 3a but for a period $\tau_{p}=100 ps$ in between the linear and the saturation regime.}
\vskip 0.5cm
In Figure 3b we increase the period to $\tau_{p}=100 ps$ in between the saturation regime (for small periods) and the linear regime (for large periods).
Due to the enlargement of time separations, the previous peaks are here replaced by two bumps, with the second bump located at the period $t=\tau_{p}$.
The first bump decreases when R increases, while the second bump increases.

\includegraphics[scale=0.34]{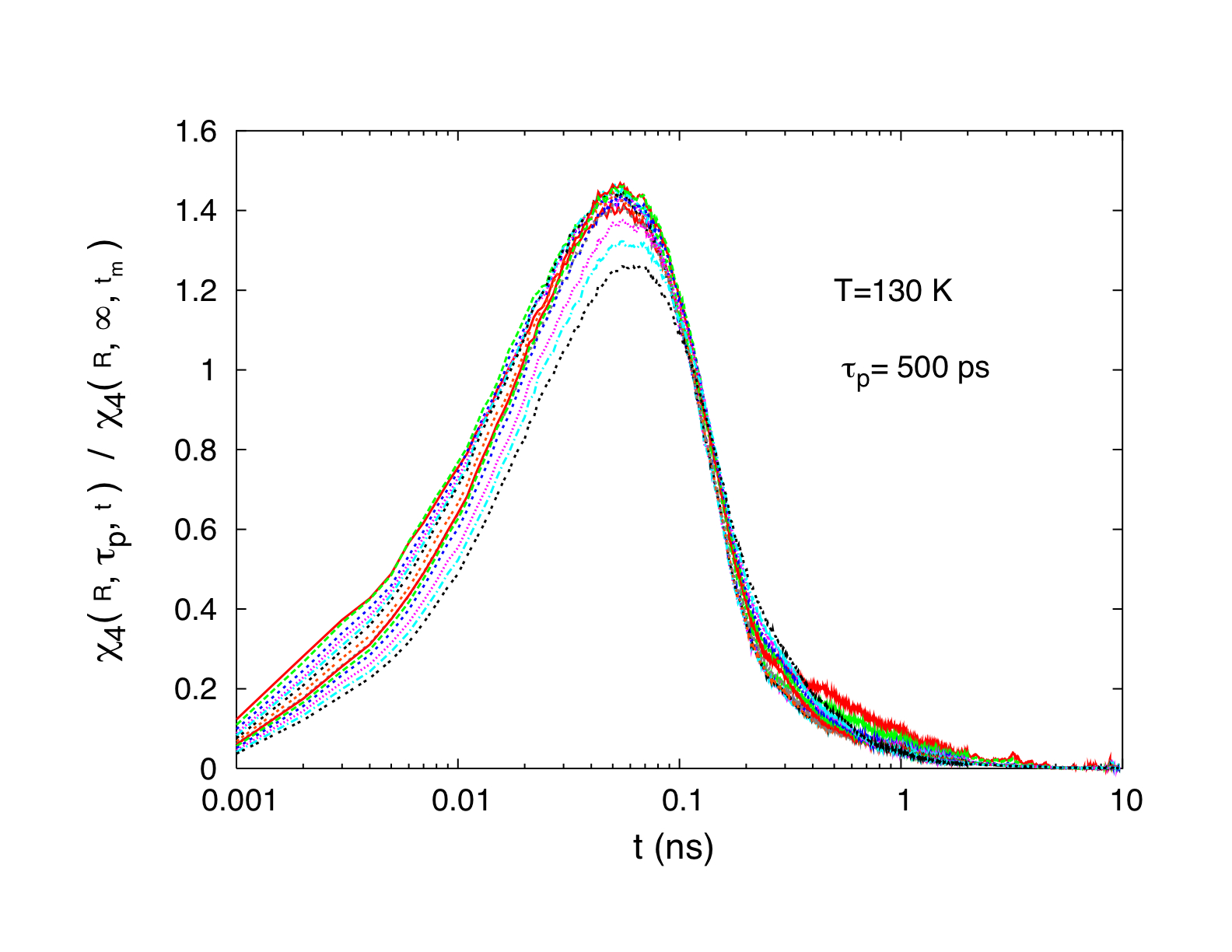}
{\em \footnotesize Figure 3c:
As in Figure 3a but for a period $\tau_{p}=500 ps$, in the linear regime .}
\vskip 0.5cm
Finally, figure 3c shows that in the linear regime the DHs only decrease more slightly when we move away from the chromophore. Moreover all the curves calculated with $R>17$ \AA\ collapse in a single curve. 
These results show that the DHs do propagate at large distance from the chromophore in the linear regime while they are strongly attenuated in the saturation regime.

\includegraphics[scale=0.34]{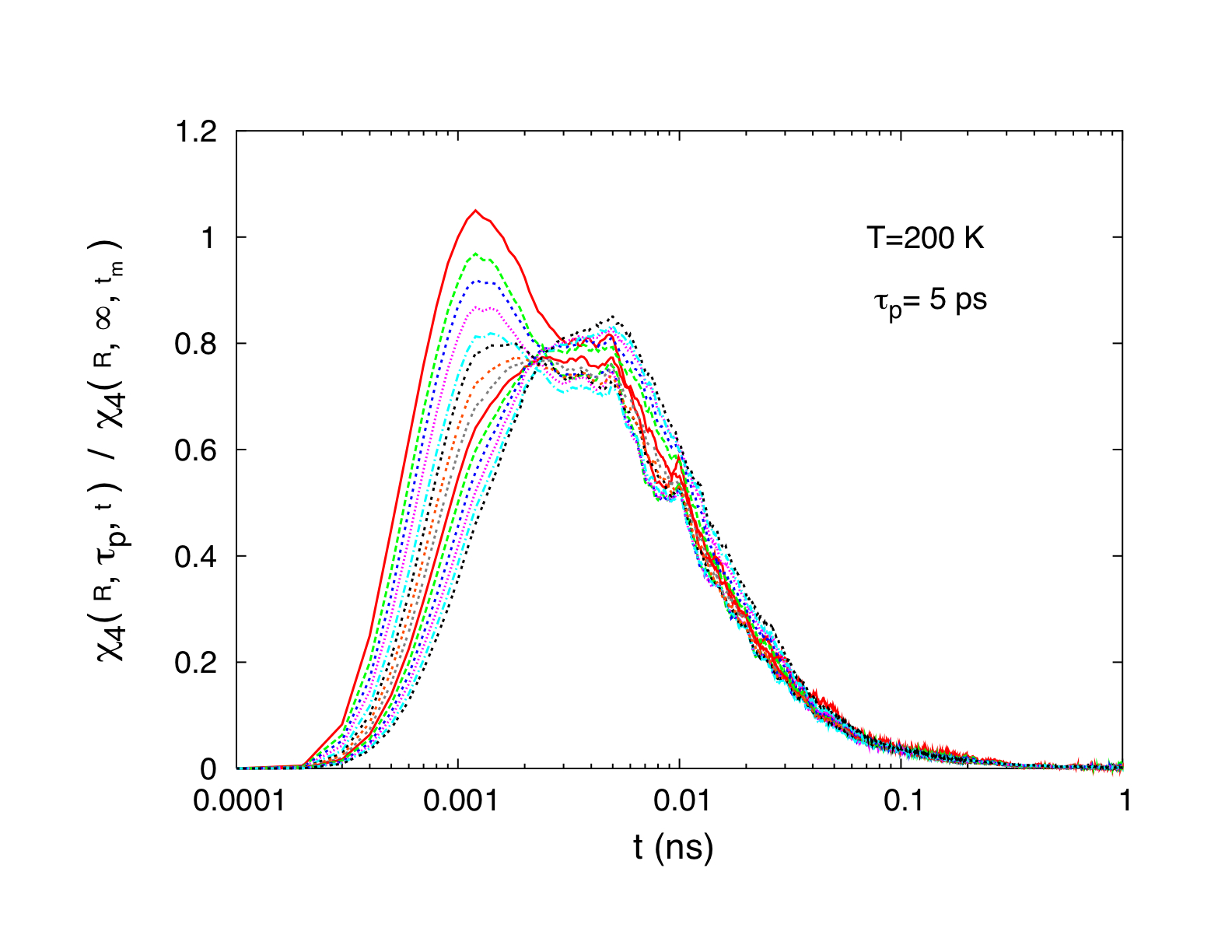}
{\em \footnotesize Figure 3d:
As in Figure 3a but for a larger temperature T=200K.}
\vskip 0.5cm
We observe the same behavior at larger temperature in Figure 3d than in Figure 3a.
However the spontaneous $\alpha$ relaxation time $\tau_{\alpha}^{(\tau_{p}=\infty)}$ of the material is shorter at that larger temperature.
As a result, the comportment observed at $\tau_{p}=5$ ps at $T=200K$ is equivalent to the comportment observed for a larger period $\tau_{p}$ at the temperature $T=130K$.

\includegraphics[scale=0.34]{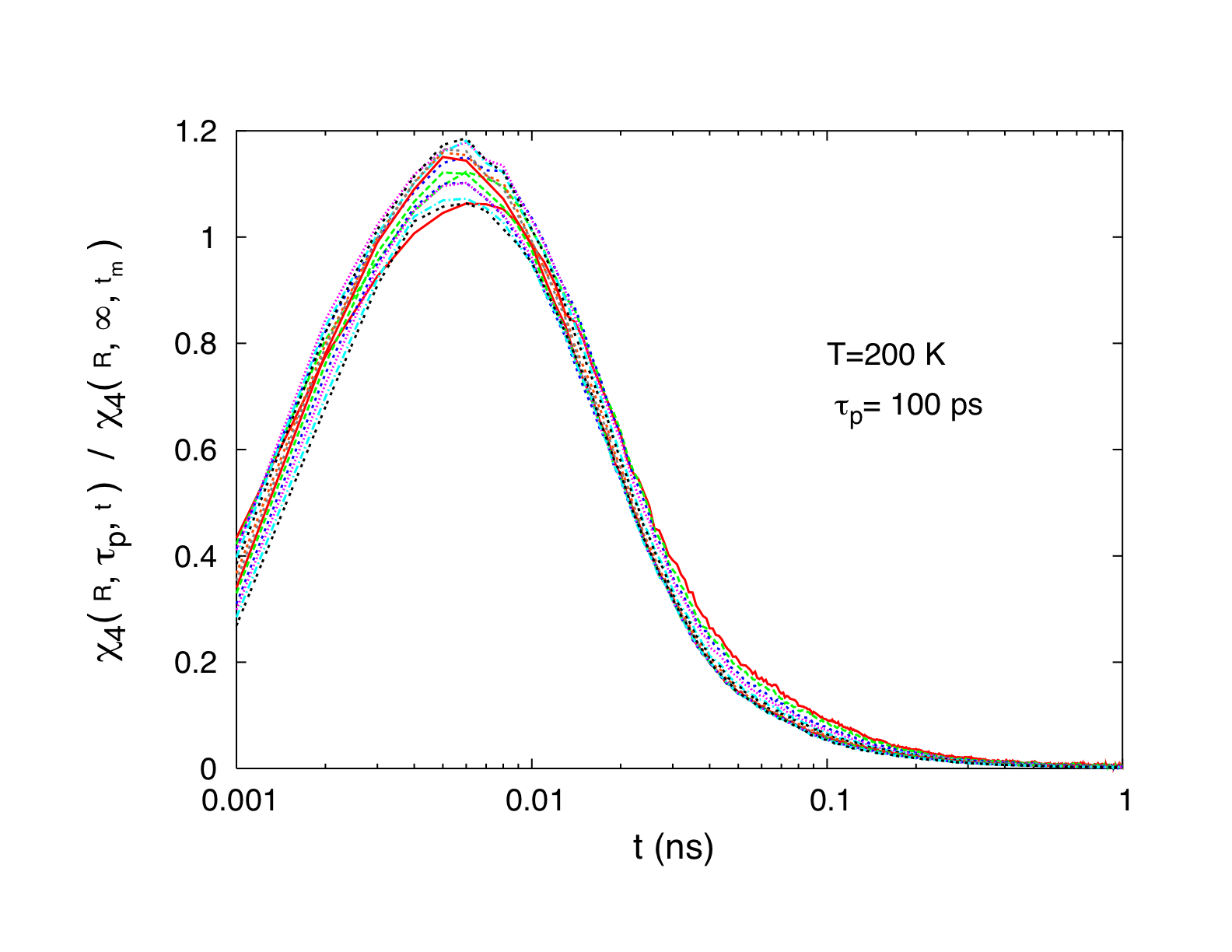}
{\em \footnotesize Figure 3e:
As in Figure 3b but for a larger temperature T=200K.
}
\vskip 0.5cm
Increasing $\tau_{p}$ we finally reach the $\alpha$ relaxation time of the material.
Figure 3e shows that in this case the heterogeneity doesn't depend anymore on the distance to the chromophore's perturbation.
The relative susceptibility is approximately equal to one, showing that we are observing mainly the spontaneous susceptibility in that Figure.

\vskip 1cm
\subsection{4. What is the evolution of the excitation concentration ?}

The DHs increase with $f$ in the linear regime, originates from the increase in the excitations concentration $C_{ex}$ with $f$. 
In order to understand better this process we will now study the evolution of the excitation concentration with the isomerization rate.
We define here excitations as molecules that move at time $t_{0}$ more than a distance $a=1.5$ \AA\ within a time lapse $\delta t=10 ps$.
 The excitation concentration is then defined as: 
\begin{equation}
C_{ex}={1 \over N.N_{t_{0}}}.  \sum_{i,t_{0}} w_{a}\left( \left\vert {{{\mathbf{r}}}}_{i}(t_{0}+\delta {t})-{{{\mathbf{r}}}}_{i}(t_{0})\right\vert \right) .  \label{e3}
\end{equation}

\includegraphics[scale=0.34]{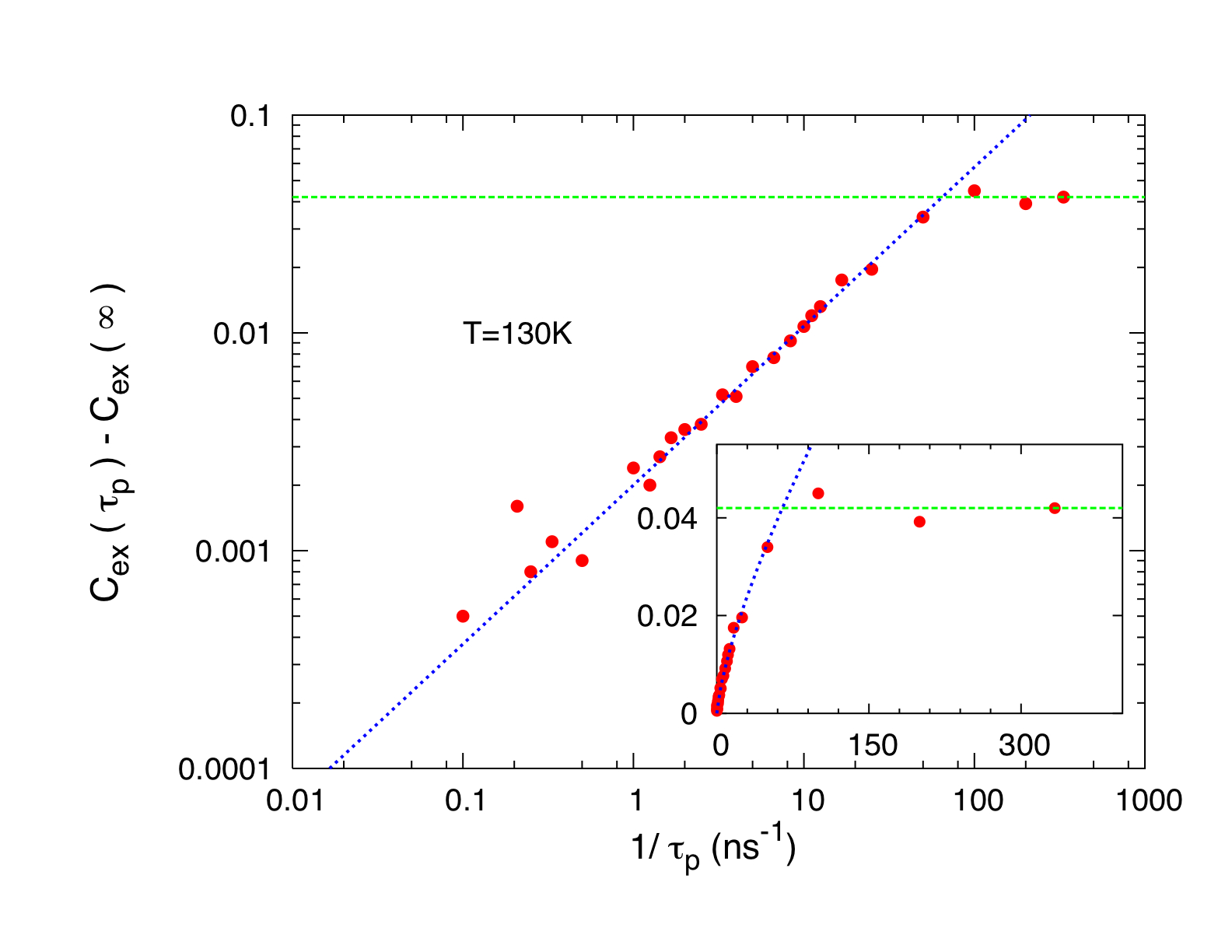}
{\em \footnotesize Figure 4a:
Isomerization-induced excitation concentration (no unit) $C_{ex}^{induced}(\tau_{p})=C_{ex}(\tau_{p})-C_{ex}(\infty)$ versus the isomerization rate $f=1/\tau_{p}$ at a temperature T=130K, in a logarithmic scale. Inset: same figure but in a linear scale.
The induced excitations concentration increases as a power law $C_{ex}^{induced}(\tau_{p})=C_{0}$ $ . (1/\tau_{p})^{0.73}$ (blue dashed line) with the isomerization rate and then saturates to a concentration $C_{ex}^{induced} \approx  5\% $ (green dashed line). 
}
\includegraphics[scale=0.34]{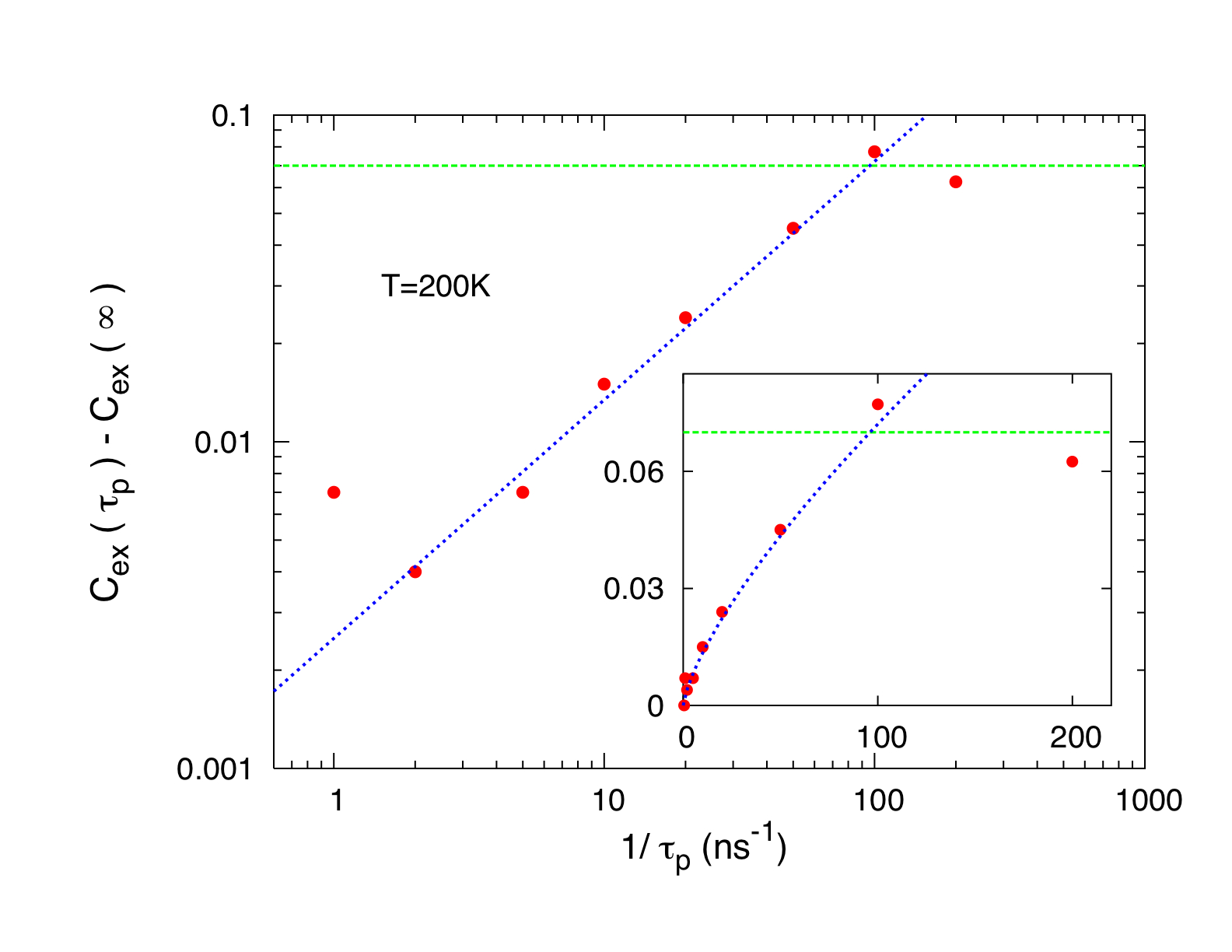}
{\em \footnotesize Figure 4b:
As in Figure 4a but at the larger temperature T=200K.
}

\vskip 0.5cm
Figure 4a and 4b show the excitation concentration evolution with the isomerization rate at two different temperatures, chosen as the lowest temperature that can be properly equilibrated without isomerization (T=130K) and the highest temperature (T=200K) for which the heterogeneities are still relatively large.
Figures 4 show that for both temperatures the induced excitation concentration $C_{ex}^{induced}(\tau_{p})=C_{ex}(\tau_{p})-C_{ex}(\infty)$ evolves as a power law of the isomerization rate $C_{ex}^{induced}(\tau_{p})=C_{0} . (1/\tau_{p})^{0.73}$ and then saturates.
This evolution is reminiscent of the evolution of the inverse of the relaxation time and of the diffusion coefficient\cite{jcpvic}, that also increases (but linearly) and then saturates.  
The evolution law is approximately the same for both temperatures with $C_{0}=0.0020$ for $T=130 K$ and $C_{0}=0.0025$ for $T=200 K$ and the same power law.
The saturation however appears for a larger period at low than at high temperature, a result explained by the increase of the relaxation time when the temperature drops.
Using the picture of excitations in a dynamically constrained system, here the dynamic constraints are created by the host material.
As a result, for small enough frequencies $f$, in the linear regime we expect the dynamics to be driven by the excitation concentration only, because the host material is only slightly affected by the isomerizations in that regime.
Thus we explain the DHs increase with $f$ in the linear regime to be due to the increase in the number of excitations.
Similarly, we expected (however the host material is in this case affected) the DHs to saturate in the saturation regime,  because the number of excitations saturates.
But instead the DHs decrease, a result that we explain from the softening of the host material in that regime.
A recent study has demonstrated \cite{hc} that in kinetically constrained models the $\alpha$ relaxation time $\tau_{\alpha}$ is connected to the excitation concentration.
In that study, the authors show that when the temperature decreases, the excitation concentration decreases, leading to larger distances between excitations and thus a smaller probability to induce the avalanches of motions that are necessary for diffusion. The authors found that as a result the $\alpha$ relaxation time is related to the excitation concentration with the power law $\tau_{\alpha} \approx C_{ex}^{-2.4}$. Our results are in qualitative agreement with that study as we also find a power law evolution of the relaxation time $\tau_{\alpha}$ with the excitation concentration. As $(1/\tau_{\alpha})^{induced}$ increases linearly with $1/\tau_{p}$ and $C_{ex}^{induced}(\tau_{p})=C_{0} . (1/\tau_{p})^{0.73}$ we find 
$\tau_{\alpha}^{induced} \approx (C_{ex}^{induced})^{\kappa}$. However our coefficient $\kappa = -1.37$ is different. We think that this difference originates from the difference in the geometry of our systems as our excitations are here concentrated around the isomerizing chromophore and not uniformly dispersed.
The similarity between these two results ($\tau_{\alpha} \approx (C_{ex})^{n}$ ) supports however the idea of a similar cause for both mechanisms, i.e. the excitation concentration.\\

\section{Conclusion}
In this paper we studied the coupling between the dynamic heterogeneities induced  by the isomerization of chromophores diluted inside a glass-former and the isomerization period of these chromophores.
To summarize our results we found that for large isomerization periods $\tau_{p}$ and small perturbations the heterogeneity increases with the isomerization rate $1/\tau_{p}$.
In contrast the heterogeneity decreases when the isomerization half-period $\tau_{p}/2$ is smaller than the relaxation time of the material $\tau_{\alpha}$. 
These two opposite behaviors appear in the same period ranges than the linear and saturation regime observed previously\cite{jcpvic} with the diffusion coefficient evolution.
Due to the increase of the DHs in the linear regime followed by a decrease in the saturation regime, the cooperative motions are maximum for $\tau_{p}/2 \approx \tau_{\alpha}$.
We also found that the heterogeneities are induced around the chromophore at a shorter range in the saturation regime than in the linear response regime, a result that explains the saturation. 
Finally we  found that the excitation concentration follows a similar evolution versus the isomerization rate than the diffusion coefficient (i.e. an increase followed by a saturation). 
Interestingly, for small enough perturbations in the picture of an unchanged medium, the increase of the diffusion coefficient could lead to the ability to probe ergodically the conformational space in shorter time ranges than without isomerizations.


\begin{thebibliography}{99}

\bibitem{h0} L. Berthier, G. Biroli, J.-P. Bouchaud, L. Cipelletti, W. Van Saarloos
 \emph{Dynamical heterogeneities in glasses, colloids and granular
media} Oxford University Press, New York (2011).



\bibitem{binder} K. Binder, W. Kob, 
\newblock  {\em  Glassy materials and disordered solids}, World Scientific, Singapore (2011)

\bibitem{bene} P.G. Debenedetti, 
\newblock  {\em  Metastable Liquids} Princeton University Press, Princeton (1996)


\bibitem{se} A. Furukawa, H. Tanaka, \emph{Phys. Rev. E} \textbf{86}, 030501(R) (2012)

\bibitem{ha} D. Chandler, J.P. Garrahan,  \emph{Annu. Rev. Phys. Chem.} \textbf{61}, 191 (2010)

\bibitem{hb} Y.S. Elmatad, A.S. Keys \emph{Phys. Rev. E} \textbf{85}, 061502 (2012)

\bibitem{hc} A.S. Keys,  L.O. Hedges, J.P. Garrahan, S.C. Glotzer, D. Chandler, \emph{Phys. Rev. X} \textbf{1}, 029901 (2011)



\bibitem{prl}  V. Teboul, M. Saiddine, J. M. Nunzi, 
\newblock  {{\em Phys. Rev. Lett. }} {\bf  103}, 265701 (2009).

\bibitem{pre}  V. Teboul, J.B. Accary, M. Chrysos, 
\newblock  {{\em Phys. Rev. E }} {\bf  87}, 032309 (2013).


\bibitem{david} D. Chandler, 
\newblock  {\em  Introduction to modern
statistical mechanics}, Oxford University Press, New York 1987

\bibitem{nature} G.J. Fang, J.E. Maclennan, Y. Yi, M.A. Glaser, M. Farrow, E. Korblova, D.M. Walba, T.E. Furtak, N.A. Clark, \emph{nature comm.} \textbf{4}, 1521 (2013)

\bibitem{natmat} P. Karageorgiev,  D. Neher, B. Schulz, B. Stiller, U. Pietsch, M. Giersig, L. Brehmer,  
{\em Nat. Mat.} {\bf 4}, 699-703 (2005).



\bibitem{cage} V. Teboul, M. Saiddine, J.M. Nunzi, J.B. Accary, 
 \newblock \emph{J. Chem. Phys. } \textbf{134}, 114517 (2011)





\bibitem{probe} M. Saiddine, V. Teboul, J.M. Nunzi, 
\newblock  {\em J. Chem. Phys.} {\bf  133}, 044902 (2010)

\bibitem{jcp} V. Teboul, J.B. Accary, 
\newblock  {\em J. Phys. Chem. B} {\bf  116}, 12621 (2012)

\bibitem{pot2} J.B. Accary, V. Teboul, \newblock  {\em J. Chem. Phys.} 
\textbf{136} 094502 (2012)

\bibitem{pot1} W.L. Jorgensen, D.S. Maxwell, J. Tirado-Rives, 
 \newblock \emph{J. Am. Chem. Soc.} \textbf{118}, 11225 (1996)




\bibitem{allen} M.P. Allen, D.J. Tildesley, 
\newblock  {\em  Computer
simulation of liquids}, Oxford University Press, New York 1990



\bibitem{beren} H.J.C. Berendsen, J.P.M. Postma, W. Van Gunsteren, A. DiNola, J.R. Haak
 \newblock  {\em  J. Chem. Phys.} {\bf 81}, 3684 (1984)

\bibitem{fermeture} C.M. Stuart, R.M. Frontiera, R.A. Mathies, 
 {\em J. Phys. Chem. A} \textbf{111}, 12072 (2007)

\bibitem{mecanisme1} Y. Ootani, K. Satoh, A. Nakayama, T. Noro, T. Taketsugu,  
\emph{J. Chem. Phys.} \textbf{131}, 194306 (2009)

\bibitem{mecanisme2} T. Fujino, S.Y. Arzhantsev, T. Tahara, 
\emph{J. Phys. Chem. A} \textbf{105}, 8123 (2001)

\bibitem{mecanisme3} G. Tiberio, L. Muccioli, R. Berardi, C. Zannoni, 
 \emph{Chem. Phys. Chem.}  \textbf{11}, 1018 (2010)


\bibitem{barrettn} T.A. Singleton, K.S. Ramsay, M.M. Barsan, I.S. Butler, C.J. Barrett, \emph{J. Phys. Chem. B} \textbf{116}, 9860 (2012)


\bibitem{jcpvic} J.B. Accary, V. Teboul,  \emph{J. Chem. Phys.} \textbf{139},
034501 (2013)

\bibitem{h5} R. Yamamoto, A. Onuki, \newblock  {\em Phys. Rev. Lett.} {\bf81}, 4915 (1998)
\bibitem{h6} R. Yamamoto, A. Onuki, \newblock  {\em Phys. Rev. E} {\bf 58}, 3515 (1998)
\bibitem{h6b} M. Hideyuki, R. Yamamoto, \newblock  {\em J. Chem. Phys.} \textbf{136}, 084505 (2012)


\bibitem{h7} H-N. Lee, K. Paeng, S.F. Swallen, M.D.  Ediger, \newblock  {\em Science} {\bf323}, 231 (2009)
\bibitem{h8} D. Orsi, L. Cristofolini, M.P. Fontana, E. Pontecorvo, C. Caronna, A. Fluerasu, F. Zontone, A. Madsen, \newblock  {\em Phys. Rev. E} {\bf 82}, 031804 (2010)










\end{thebibliography}
\end{document}